# Modelling and Simulation of of high efficiency GaAs PIN-Solar Cell

Ali Imran[a,*], Deborah Eric[b], Muhammad Noaman Zahid[b], Muhammad Yousaf[c],
[a]State Key Laboratory for Artificial Microstructures and Mesoscopic Physics, Peking University, Beijing
[b]School of Optics and Photonics, Beijing Institute of Technology, Beijing
[c]Department of Material Science and Engineering, Peking University, Beijing
E-mail: aliimran@pku.edu.cn

**ABSTRACT**

Solar energy is the most convenient and reliable energy source among all renewable energy resources and an efficient photovoltaic device is required to convert this energy into utilizable energy. Different types of solar cells (SC) are commercially available. However, various parameters need to be optimized to get maximum efficiency from a SC. In this study we have presented a SC model in which dependence of quantum efficiency (QE) on various parameters has been investigated. The mobility of the carriers has been varied with wide range along with the carrier life time (LT). Results show that maximum efficiencies can be achieved up to 11.10% and 10.81% keeping the electron and hole mobility to be 1500 $cm^2V^{-1}s^{-1}$ and 300 $cm^2V^{-1}s^{-1}$ respectively with electron and hole carrier LT to be 3ns and 7ns respectively. The effect of surface recombination velocity (SRV) has also been brought under observation and the maximum efficiency is found to be 13.75% at electron and hole SRV equal to be $10^3 ms^{-1}$. Results shows that the higher photovoltaic efficiencies can be achieved by increasing the mobility and carrier LT while decreasing the surface recombination velocities.

## 1. INTRODUCTION

Conversion of solar energy into electric energy is a remarkable renewable alternative for the electricity generation. Various other technologies have also been used to harvest energy from natural resources such as wind, water and biofuels [1-5]. Among these all, solar energy is the most promising alternative. The discovery of Photovoltaic technology has made a breakthrough in the modern human life, not only renewable but also clean and environment friendly. Scientific communities around the globe are working on increasing the efficiency and decreasing the cost from many decades. The SC available in the commercial market is still made by silicon. Recently, III-V compounds semiconductors are used in optoelectronics and SCs with higher efficiencies compared to silicon SCs [6-9]. Among these compound semiconductors, GaAs is currently one of the most attractive candidates for photovoltaic applications due to its lower temperature coefficient and higher electron mobility and higher density of states compared to silicon SCs [10-15]. The surface recombination (SR) has a great effect on the performance of a SC, decreasing its photocurrent and increasing dark current, particularly for small area cells. SR is reported to be the cause of an efficiency loss of 4% in SCs [16-19].

The first model of GaAs SC was heterostructure PN junction and it was discovered that the decrease in the output current is much affected due to SR compared to recombination process inside junction region. So the analytical expression of this model was made which was based on the fundamental assumptions which includes Fermi level pinning and doping density of electron and hole [20]. The SR current was found to decrease away from the junction according to surface diffusion length (DL). Later recombination was treated at the PN junction surface similar to inside the depletion layer width having SR current with an effective width (EW) [21, 22] . The EW is a very small fraction of the junction depletion layer width. It was observed that electric field at the surface has the same direction but with a much lower value due to the presence of charged surface states. The diffusion of minority carriers along the junction surface was neglected. The calculated EW value was found to be much smaller than the surface DL reported in early studies [23-28]. It was discovered that the SR current is due to the drift diffusion of carriers inside the depletion region and this current is dominant at low bias. At higher voltages it becomes comparable to recombination current resulting from carrier injection from outside the junction. [29-32].

The Numerical simulation is the best tool for optimizing the structures of photovoltaic devices due to cost, time and device fabrication complexity reduction. It is one of the best ways in recent years to design a SC. Precise modeling is used to understand internal physics of SC with different assumptions to define analytical descriptions. [33-38]. In this research work, we have investigated the variation in QE effected by carrier mobility, SRV and carrier LT. All of the simulations are performed on COMSOL Multiphysics software.

## 2. THEORY AND METHODS

The photovoltaic device in this research consists of a P-layer, I-layer and N-layer as shown in Figure 1. The widths of the P, I and N layer are taken to be 200 nm, 600 nm and 100nm respectively. The doping concentrations of P-layer and N-layer are taken to be $10^{17}$ cm$^{-3}$ while the I-layer is not doped. We have investigated the results using different values of mobility, carrier LT and surface recombination. The parametric values used are taken from experimental data [39]. The simulation is performed by COMSOL Multiphysics. Following are the equations involved in this Multiphysics calculations. The optically generated current is directly proportional to the incident photon flux. The number of incident photons can be calculated on the surface of the SC from incident power.

$$F_0 = \frac{P_{in}}{E} \tag{1}$$

Where E is the energy of the photon which depends on the frequency.

$$E = h\nu \tag{2}$$

The transverse electromagnetic mode of the incident electric field for the SC surface along.

$$E = E_0 e^{-ikx} \tag{3}$$

The poison's equation is applied to the calculation of the electric field generation, which depends on carrier densities.

$$\frac{\partial E}{\partial x} = \frac{\rho}{\varepsilon} \tag{4}$$

Where $\rho$ is the charge density.

$$\rho = q(-n + p + N_d - N_a) \tag{5}$$

Where $N_a$ and $N_d$ is the donor and acceptor concentration. The total drift and diffusion current can be calculated by

$$J = J_{drift} + J_{diffusion} \tag{6}$$

Where

$$J_n = q\rho_n \mu_n E + qD_n \frac{\partial n}{\partial r} \tag{7}$$

$$J_p = q\rho_p \mu_p E + qD_p \frac{\partial p}{\partial r} \tag{8}$$

Where $\mu_n$ and $\mu_p$ is the mobility of electron and hole respectively. $D_n$ and $D_p$ are diffusion coefficient of electron and hole respectively. The output current can be found as

$$J_{out} = (G - U) \tag{9}$$

Where the generation (G) can be found as

$$G = \alpha F_0 e^{-\alpha z} \tag{10}$$

Where α is absorption coefficient, $F_0$ is photon flux at the surface per unit area, and z is the penetration depth into the material. We have considered trap assisted Shockley Read Hall and surface recombination which can be given as

$$U_{SRH} = \frac{np - n_i^2}{\tau_p(n+n_0) + \tau_n(p+p_0)} \tag{11}$$

Where $n$ and $p$ are the numbers of electrons and holes, $n_0$ and $p_0$ are the number of electrons and holes in equilibrium. The carrier LTs in band to band recombination can be found as.

$$\tau = \frac{C}{N_{DOPING}} \tag{12}$$

Where $C$ is the constant and its value for direct band gap materials is $10^{10}$ s.cm$^{-3}$ [39, 40]. SR for electrons and holes are

$$U_{S_p}\delta z = S_n(n_s - n_0) \tag{13}$$

$$U_{S_n}\delta z = S_p(p_s - p_0) \tag{14}$$

Whereas while $n_0$ and $p_0$ are the number of electrons and holes in equilibrium, while $n_s$ and $p_s$ are the electron and hole densities at the surface

$$n_s = n_p = n_i \exp\left(\frac{eV}{kT}\right) \tag{15}$$

The SRV for electrons and holes are $S_n$ and $S_p$ respectively and can be given by,

$$S_n = \upsilon_n \sigma_n N_n \tag{16}$$

$$S_p = \upsilon_p \sigma_p N_p \tag{17}$$

$N_n = N_p$ ($10^{11}$cm$^{-2}$) surface traps at energy level $E_s$ (which is taken to be energy at intrinsic level) per unit area within a layer $\delta z$, $v_n$ and $v_p$ are the thermal velocities of the electron and hole, respectively, $\sigma_n = 1.4 \times 10^{-16}$ cm$^2$ and $\sigma_p = 1.4 \times 10^{-18}$ cm$^2$ are the capture cross section of the trap for electrons and holes [41-43]. The total current density can be found as,

$$J_D = J_n + J_p \tag{18}$$

The output voltage for this case will be

$$V_{OC} = \frac{nkT}{q} \ln\left(\frac{J_L}{J_D} + 1\right) \tag{19}$$

Efficiency can be found as,

$$\eta = \frac{P_{out}}{P_{in}} = \frac{V_m I_m}{P_{in}} \tag{20}$$

## 3. RESULTS AND DISCUSSIONS

The aim is to study the effect of carrier mobilities, LTs and SRV on the QE of the SC. The solar spectrum used in this study is the AM 1.5 spectrum with total power concentration of 100Wm$^{-2}$. Light propagation inside the device is investigated by applying Fresnel equation and corresponding electrical field is calculated. This is followed by simulation of electrostatic drift diffusion model of the SC in the dark situation. The carrier concentration at equilibrium is also calculated, while the built-in voltage is calculated by solving Poisson equation. The whole model is simulated under illumination by applying dynamic electrostatic equilibrium condition. We have applied finite element methods to solve Maxwell's equations. There are four regions (Air, N-layer, I-layer and P-layer) and the propagation equation for each region is solved as shown in Figure.1 (a) Solar radiation is assumed to be at normal incidence, where a scattering boundary condition is used for the top surface and perfect electric conductor boundary condition for the bottom metallic refractor. Figure.1 presents the results for three different frequencies for 530nm, 730nm and 930nm.

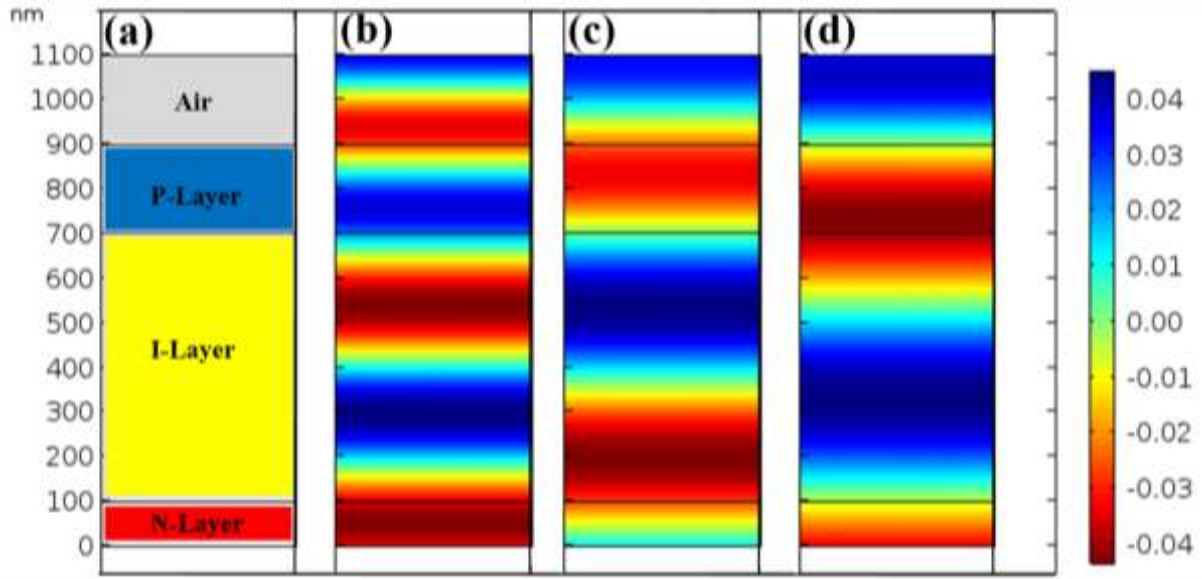

Figure. 1(a) Schematic diagram of PIN solar cell and electrical filed at the wavelength of (b) 550nm(c) 650nm and (d) 750nm

The built-in potential in device depends on the doping concentration of donors and acceptors. The solution of our simulation gives rise to the electrostatic potential of the device, based on the distribution of carriers, which gives an energy band diagram at equilibrium with no currents owing at the contacts. The blue color shows the concentration of excess of electrons while red shows the holes. The green color is the intrinsic region with no doping. The graphical doping profile of the semiconductor is presented in Figure. 2 having the maximum donor and acceptor concentration of the order $10^{21}$ m$^{-3}$.

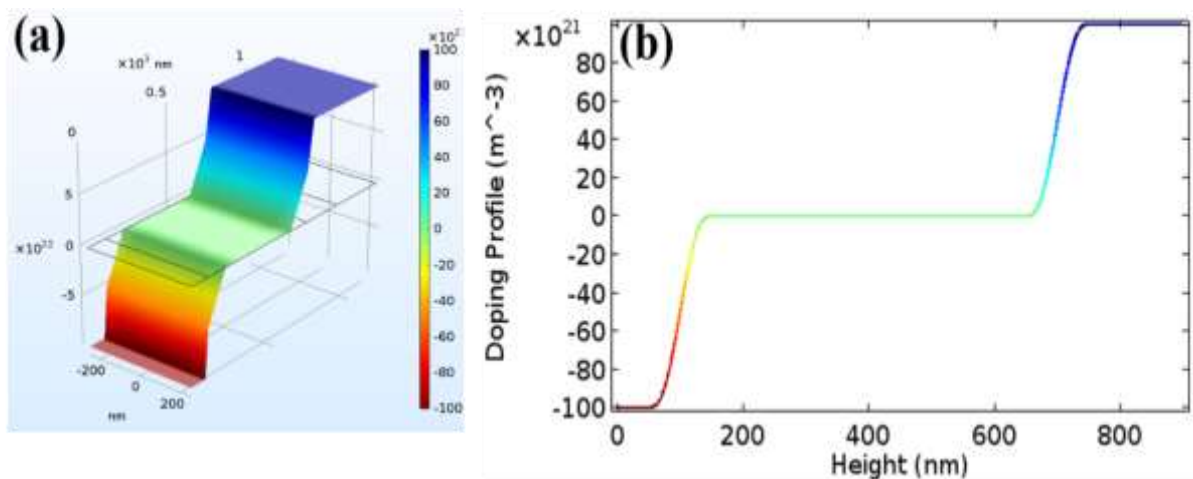

Figure. 2(a) Three dimensional doping profile PIN solar cell (b) Linear doping profile PIN solar cell

The variation in the overall space charge densities under different applied voltages are investigated, which coupled to the Poisson equation along with electrons and holes continuity equations through Multiphysics coupling available in COMSOL. Figure 3(b) shows that there is no charge density accumulation in the intrinsic region of the device while N and P regions have very high values of carrier concentrations.

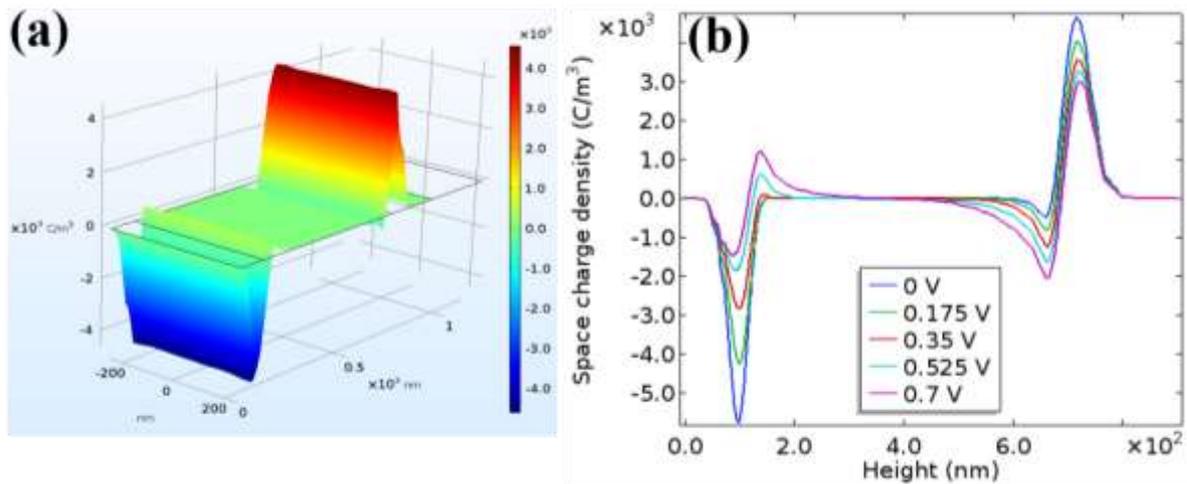

Figure. 3(a) Three dimensional space charge density profile and (b) Space charge density graphical profile for different voltage

The charge density is further used to calculate the electron and holes concentrations under illumination through drift diffusion model. The concentrations of free charge carries are presented for different values of voltage in Figure. 4(b) and 4(d). The solution gives carrier concentrations coupled to the electrostatic potential, which are used to compute the total current densities throughout the full structure using drift and diffusion model. The carrier concentration increased as the device is exposed to radiation, so is the effect while the device is subjected to bias to calculate I-V characteristics.

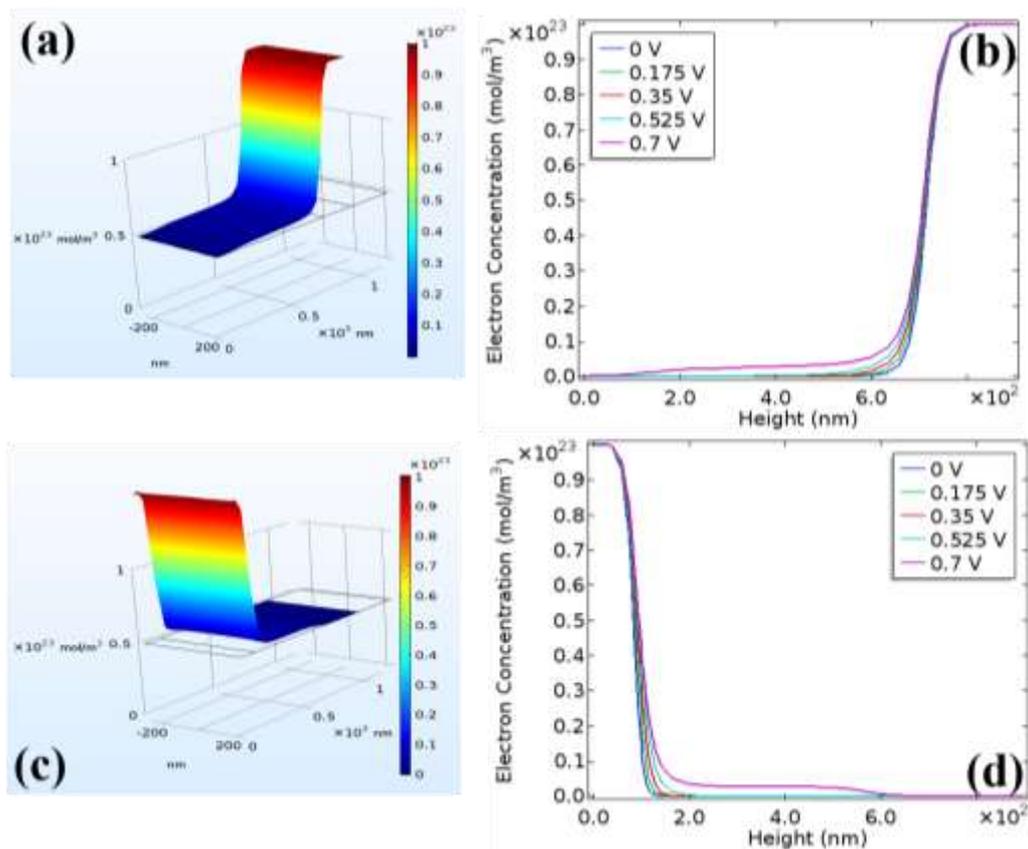

Figure. 4(a) Three dimension electron concentration profile (b) electron concentration profile for different voltage(c) Three dimension hole concentration profile (d) hole concentration profile for different voltage

## 3.1 The effect of varying the mobility

The mobilities of the electrons in the P-layer and the holes in the N-layer are given in Table 1. The effect on the QE by varying the electron and hole mobility is shown in Figure 5(a) and Figure 5(b). The effect on the I-V characteristic by varying both of the mobilities is shown in Figure 5(c). The lowest mobility in this calculation is 1000 $cm^2V^{-1}s^{-1}$ which gives a maximum QE of 0.44, while the highest mobility of 1500 $cm^2V^{-1}s^{-1}$ gives a maximum of 0.46. It can be seen the difference in QE is less for longer wavelengths. The reason for this is that for long wavelengths the absorption coefficient is small and has a larger effect on the QE than the mobility. The mobilities of 125 $cm^2V^{-1}s^{-1}$ and 12500 $cm^2V^{-1}s^{-1}$ are included to study the effect of varying the mobility. It can be seen the QE obtained for the mobility of 125 $cm^2V^{-1}s^{-1}$ is much smaller for all wavelengths than the QE obtained for the mobility of 12500 $cm^2V^{-1}s^{-1}$. This gives a much smaller photocurrent.

The effect on the QE of the N-layer by varying the mobility of the holes is shown in Figure 5(b), where it can be seen that it has not the same significance as varying the mobility in the P-layer. Since much of the photons are absorbed in the preceding layers, the QE in the N-layer is in any case low. The lowest value of mobility in this calculation gives a maximum of QE of 0.107 while the highest mobility a maximum of QE of 0.109. By increasing the mobility from 27.8 $cm^2V^{-1}s^{-1}$ to 278 $cm^2V^{-1}s^{-1}$, the maximum of the QE increases from 0.108 to 0.146. While the relative change might be comparable with the change of QE in the P-layer by varying the mobility, the corresponding change in total photocurrent is small since the N-layer quantum efficiencies are small. The conclusion is that a high mobility of electrons in the P-layer is more important than the mobility of the holes in the N-layer. In contrast to the case for the P-layer, the effect of varying the mobility in the N-layer increases for longer wavelengths. This is because the absorption in the N-layer is larger for the longest wavelengths and hence only a small part of these photons is absorbed in the preceding layers. Thus the importance of the hole mobility of the N-layer increases with wavelength. In Figure 5(c) the I-V characteristics of the PIN SC for various mobilities are shown. The photocurrent increases with increasing mobilities since this increase was seen for the quantum efficiencies. The open circuit voltage is however reduced with increasing mobilities, as the dark current increases. In Table. 1, the short-circuit current, the dark current, the open-circuit voltage and the resulting efficiencies for various mobilities is given. The effect of varying the mobilities can be seen from the Table. 1 which indicate that the efficiency enhances with increasing mobilities.

**Table. 1 Effect of carrier motilities on the efficiency**

| $\mu_n(cm^2V^{-1}s^{-1})$ | $\mu_p(cm^2V^{-1}s^{-1})$ | $J_{sc}(Am^{-2})$ | $V_{oc}(V)$ | $\eta$ (%) |
|---|---|---|---|---|
| 125 | 27.8 | 92.10 | 0.98 | 7.44 |
| 1000 | 250 | 128.20 | 0.96 | 10.43 |
| 1250 | 278 | 132.70 | 0.96 | 10.80 |
| 1500 | 300 | 136.20 | 0.95 | 11.10 |
| 12500 | 2780 | 165.8 | 0.94 | 13.41 |

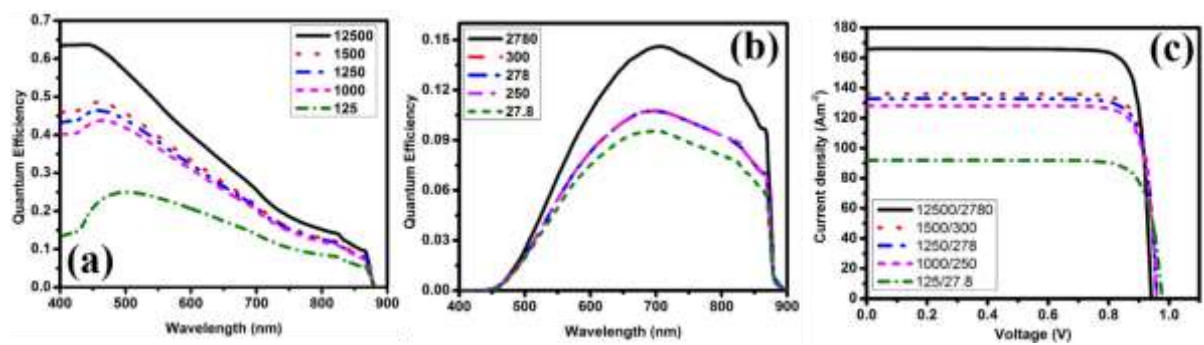

Figure .5(a) QE using different values for the electron mobility $\mu_n$ ($cm^2V^{-1}s^{-1}$) in P-layer and (b) QE using different values for the hole mobility $\mu_p$ ($cm^2V^{-1}s^{-1}$) in N-layer (c) I-V characteristic of solar cell for various ($\mu_n$ /$\mu_p$) values.

## 3.3 The effect of varying the lifetimes in the doped layers

The variation of LTs in the doped layers also effect on the efficiency of SC. The minority carrier LTs for electrons in the P-layer and holes in the N-layer are given in Table 2. The effect of varying the lifetimes in the P-layer and in the N-layer is shown in Figure 6. The LTs had to be varied this much to see an effect in the QE. It can be seen that there is no any obvious change take place in QE with increasing the LTs by a factor of $10^3$, however decreasing the lifetimes by same factor there is an apparent change in QE on both the P and N-layer.

**Table. 2: Efficiency variation due to carrier life times.**

| $\tau_n$(ns) | $\tau_p$(ns) | $J_{sc}$(Am$^{-2}$) | $V_{oc}$(V) | $\eta$ (%) |
|---|---|---|---|---|
| 3x10$^{-3}$ | 7x10$^{-3}$ | 79.5 | 0.92 | 6.23 |
| 3x10$^0$ | 7x10$^0$ | 132.7 | 0.96 | 10.80 |
| 3x10$^3$ | 7x10$^3$ | 132.8 | 0.96 | 10.81 |

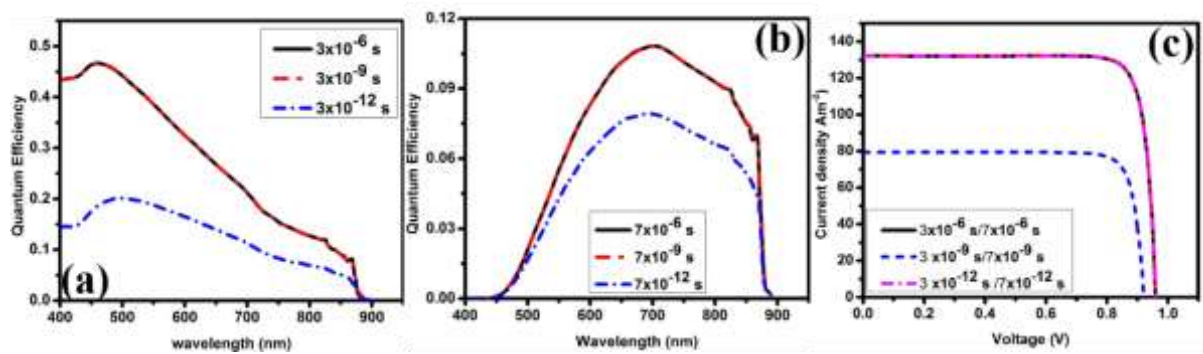

Figure. 6(a) QE using different values for the electron lifetime $\tau_n$ (ns) in P-layer and (b) QE using different values for the hole lifetime $\tau_p$ (ns) in N-layer (c) I-V characteristic of solar cell for various electron/hole lifetimes ($\tau_n/\tau_p$).

The reason for this high value of efficiency is that by using electron and hole LT 3 ns and 7 ns respectively, the values for the DLs for electrons and holes are 3114 nm and 4833 nm respectively. These values of DLs are larger than the widths of the P and N-layer. The actual value of the DL is not that important when it is larger than the P and N-layer widths. By decreasing the LTs in the P and N-layer by a factor $10^3$ the DLs are less than the widths of the P and N-layer and thus the QE dependency of the DLs is calculated which can be seen in Figure 6. When the P-layer is much thicker than the DL, the QE is reduced. In that case part of the P-layer absorbs light, but does not contribute to the photocurrent. The reduction of QE in the N-layer by decreasing the LT can be seen in Figure 6(b). A very small change in photocurrent by changing the LT in the N-layer is observed compared with P-layer. This is because the N-layer is placed on the bottom of the SC where most of the photons are already absorbed.

The effect on the I-V characteristics by varying the LTs is shown in Figure 6(c). The same behavior can be observed in the I-V characteristics as in the QE curves. By increasing the LTs, the change in the current voltage curve is not seen, while decreasing the LT by a factor $10^3$, a remarkable change in the I-V curve is observed. The reason for this is the same as mentioned for the quantum efficiencies, when the DLs are much longer than the widths of the P and N-layer they do not affect the dark-current much. When they are shorter than the widths of the P and N-layer they clearly affect the dark-current. For DLs shorter than the widths of the layers, the dark-current increases with decreasing LTs, and as a result the open-circuit voltage reduces. The open-circuit voltage, photocurrent and efficiency for the various values of the electron and hole LTs are listed in Table. 2. The conclusion drawn from the I-V and QE curves is that the uncertainties in the LTs for electrons and holes are not important in the modeling, as long as the width of the P and N-layer are smaller than the corresponding DLs.

## 3.3 The effect of varying the surface recombination velocities

Along with the effect of carrier motilities and LT on voltage-current characteristics, the SRV is also an important parameter for the account of QE. The SRV at the edge of the P and N-layer are taken to be $10^4$ ms$^{-1}$ for both carriers. The effect on the QE by varying the SRV by at the edge of the P-layer is shown in Figure 7(a) and 7(b) respectively. Maximum 0.64 QE in the P-layer is observed for $10^3$ ms$^{-1}$ electron SRV which decreases to 0.47 at $10^4$ ms$^{-1}$. It shows that the value of the SRV is very important parameter in determining the photocurrent. It is also observed that the value of the SRV in the P-layer is most important, especially for the short wavelengths where the absorption near the surface is highest. This is the same behavior as the mobility. Similarly, the behavior of the QE of the N-layer shows that the SRV here is more sensitive for longer wavelength as absorption is observed to be highest in this case. As for the mobilities and the LTs, the value of the SRV in the P-layer is more important than in the N-layer since the QE is larger. The dark current increases with increasing SRV resulting in lower open-circuit voltages for higher SRV, as shown in Figure 7(c). A high SRV gives both a low photocurrent and a low open-circuit voltage which both lead to a low efficiency. The photocurrents, open circuit voltages and efficiencies for various recombination velocities are listed in Table. 3. The front SRV is the most important, however this can be reduced using a heavily doped front surface layer.

**Table. 3: Efficiency variation due to surface recombination velocities.**

| $S_n$(ms$^{-1}$) | $S_p$(ms$^{-1}$) | $J_{sc}$(Am$^{-2}$) | $V_{oc}$(V) | η (%) |
|---|---|---|---|---|
| 1x10$^3$ | 1x10$^3$ | 165.6 | 0.99 | 13.75 |
| 1x10$^4$ | 1x10$^4$ | 132.7 | 0.96 | 10.80 |
| 1x10$^5$ | 1x10$^5$ | 92.4 | 0.94 | 7.37 |

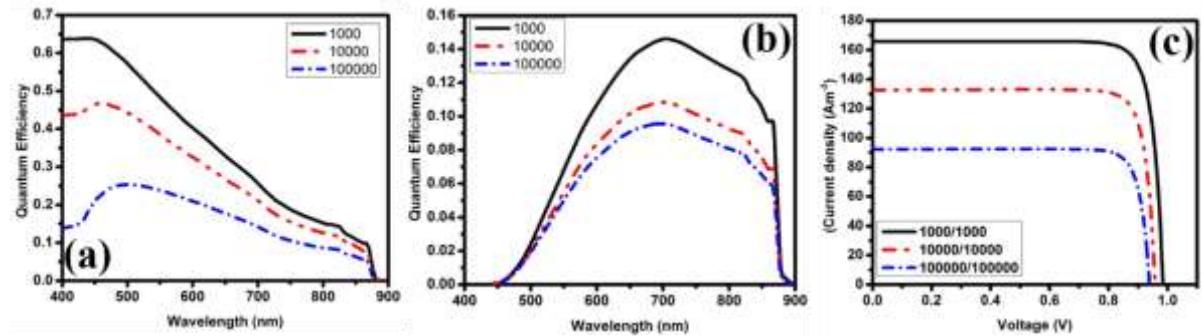

Figure .7(a) QE using different values of $S_n$ (ms$^{-1}$) in P-layer (b) QE using different values of $S_p$ (ms$^{-1}$) in N-layer (c) IV characteristic of solar cell for various ($S_n$/$S_p$) values.

## 4. CONCLUSION

In this study, various important parameters used for the modeling of SCs have been varied and their effect on the QE and I-V characteristic has been presented. It has been investigated that the most important parameters in determining the QE are the mobility of the electrons and the SRV in the P-layer which is the front surface. A high mobility and a low SRV give a large photocurrent. Variations in their magnitudes greatly affect the photocurrent, the open circuit voltage and efficiency of the SC. The maximum efficiencies can be achieved up to 11.10% at the electron and hole mobility equal to be 1500 cm$^2$V$^{-1}$s$^{-1}$ and 300 cm$^2$V$^{-1}$s$^{-1}$. The mobility and SRV also affect the dark current together with the carrier LT. The LT of the minority carriers in the doped layers are not important if the width of the layers is smaller than the DLs. The efficiency up to 10.81% is observed with electron and hole carrier LT to be 3ns and 7ns respectively. High output values can be obtained at the lower SRV values and the maximum efficiency is found to be 13.75% at $10^3$ms$^{-1}$ electron and hole SRV. It is concluded that best results can be achieved at minimum recombination velocity, high carrier LT and high values of mobilities.